\documentclass[prb,twocolumn,showpacs,floatfix]{revtex4}
\usepackage{epsfig}
\usepackage{amssymb}
\usepackage{amsmath}

\begin{document}

\title{Local Ferromagnetism in Microporous Carbon with the Structural Regularity of Zeolite Y}

\author{Y. Kopelevich}
\author{R. R. da Silva}
\author{J. H. S. Torres}
\author{A. Penicaud}
\altaffiliation{On leave from CNRS, UPR 8641, Universit\'{e} Bordeaux-I, av. Schweitzer, 33600, Pessac, France.}
\affiliation{Instituto de F\'{\i}sica ``Gleb Wataghin'', 
Universidade Estadual de Campinas, Unicamp 13083-970, Campinas, S\~{a}o Paulo, Brasil}
\author{T. Kyotani}
\affiliation{Institute of Multidisciplinary Research for Advanced Materials, Tohoku University, 2-1-1 Katahira,
Aoba-Ku, Sendai 980-8577, Japan}

\begin{abstract}

Magnetization M(H,T) measurements have been performed on microporous carbon (MC) 
with a three-dimensional nano-array structure corresponding to that of a zeolite 
Y supercage. The obtained results unambiguously demonstrate the occurrence of  
high-temperature ferromagnetism in MC, probably originating from a topological 
disorder associated with curved graphene sheets. The results provide evidence that 
the ferromagnetic behavior of MC is governed by isolated clusters in a broad temperature range, 
and suggest the occurrence of percolative-type transition with the temperature 
lowering. A comparative analysis of the results obtained on MC and related materials is given.

\end{abstract}

\maketitle

During the last approximately 15 years, various research groups have reported on 
the occurrence of room-temperature ferromagnetism in materials made solely of 
carbon. Thus, high-temperature ferromagnetic (FM) behavior has been observed in pyrolitic 
carbon \cite{1}, amorphous-like carbon prepared by direct pyrolysis \cite{2}, 
super-high surface area carbon with micro-graphitic structures \cite{3,4}, 
C$_{60}$ under photoassisted oxidation \cite{5}, highly oriented pyrolitic graphite 
(HOPG) \cite{6,7}, extraterrestrial graphite \cite{8}, and polymerized 
rhombohedral (rh) C$_{60}$ \cite{9,10,11}. Certainly, an understanding of 
mechanism(s) behind the magnetic behavior of the carbon-based materials as well 
as an engineering of novel FM carbon structures should have a wide impact.

It appears that the largest among reported low-temperature values of the 
spontaneous magnetization M$_{s}$ ($\sim 0.05 \ldots 0.5 emu/g$ \cite{12}) 
take place in nearly graphitized materials which contain amorphous and/or 
fullerene-like carbon fragments \cite{1,9,10,11}. Indeed, the high-temperature 
FM in rh-C$_{60}$ compounds \cite{9,10,11} occurs only in samples prepared 
very close to the temperature at which fullerene cages collapse and an 
amorphous carbon forms. The presence of amorphous carbon in nominally 
rh-C$_{60}$ and its possible decisive role in the magnetic behavior of this 
material has been pointed out in Ref. \cite{11}. On the other hand, experiments 
performed on a glassy carbon (GC) \cite{13} indicate that ferromagnetism emerges during 
the graphitization process, and a possibility that topological defects trigger 
ferromagnetism has been suggested \cite{13}. In fact, the occurrence of FM, 
antiferromagnetic (AFM), and superconducting (SC) instabilities due to a 
topological disorder in graphitic sheets has been 
predicted theoretically \cite{14}. The analysis given in Ref. \cite{14} assumes the 
formation of pentagons and heptagons, i. e. disclinations in the graphene 
honeycomb lattice. The low-lying electronic states of an isolated graphitic 
sheet can be well approximated by Dirac equations in (2+1) dimensions. 
Then, according to Ref. \cite{14}, a random distribution of topological 
defects described in terms of a random gauge field can lead to an enhancement 
of the density of states of Dirac fermions N(E) at low energies E, and hence 
to magnetic or SC instabilities. Compounds with curved graphene layers 
have been proposed as promising materials for both FM and SC occurrence \cite{14}.

The aforesaid experimental as well as theoretical results motivate us to explore 
magnetic behavior of the microporous carbon (MC) with a three-dimensional (3D) 
nano-array structure whose arrangement matches that of supercages of zeolite Y. 
Details of the MC samples preparation and their characterization are given 
elsewhere \cite{15,16}. Briefly, the MC has been prepared by the following template
technique. Powder zeolite Y was impregnated with furfuryl alcohol (FA) and FA was
polimerized inside the zeolite channels by heating the FA/zeolite composite at
150$^{\circ}$C under N$_{2}$ flow. The resultant polyfurfuryl alcohol/zeolite
composite was heated to 700$^{\circ}$C in N$_{2}$. As soon as the temperature reached
700$^{\circ}$C, propylene chemical vapor deposition (CVD) was performed 
for further carbon deposition. After the CVD, the composite was further
heat-treated at 900$^{\circ}$C under a N$_{2}$ flow. The resultant carbon was liberated
from the zeolite framework by acid washing. The obtained MC possesses a very high 
surface area of 3600 m$^{2}$/g and consists of a curved 3D graphene network which 
may contain randomly distributed pentagons and heptagons. The carbon particle size 
ranges from 1000 \AA~to 4000 \AA \cite{16}. Spectrographic analysis of the MC samples 
reveals magnetic impurity contents of Fe (64 ppm), Co (4.4 ppm), and Ni (3.5 ppm). 

We performed dc magnetization M(H,T) measurements on two MC samples from the same 
batch with masses m$_{1}$ = 10.82 mg (labeled as MC1) and m$_{2}$ = 8.7 mg (labeled as 
MC2) in the temperature range 2 K $\leq$ T $\leq$ 300 K and applied magnetic field up 
to $\mu_0$H = 5 T using the SQUID magnetometer MPMS5 (Quantum Design). 

The obtained results unambiguously demonstrate the occurrence of FM in our MC samples. 
The results also provide evidence that the sample FM behavior is governed by isolated 
or weakly interacting clusters (droplets) in a broad temperature range 
T$_{c}$ $\leq$ T $\leq$ T$^{*}$, where T$^{*} > 300$ K and T$_{c}$ $\approx$ 30 K. 
Below T$_{c}$, FM magnetization suddenly increases suggesting a percolative-type 
transition associated with growing in size and/or interacting FM clusters. 
At yet lower temperature T $\leq$ T$_{\times}$ $\approx$ 10 K, a magnetization anomaly 
seemingly associated with competing AFM order has been observed.

\begin{figure}
\begin{center}
\epsfig{file=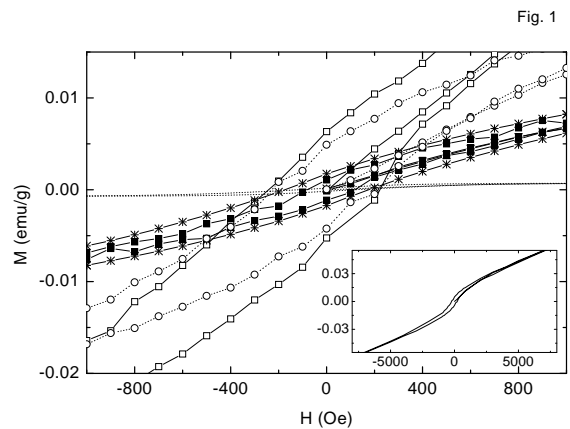,width=\columnwidth}
\end{center}
\vspace{-1cm} \caption[*]{Low-field portions of M(H) obtained for MC1 sample in 
the field range -50 kOe $\leq$ H $\leq$ 50 kOe and T = 2 K ($\square$), 5 K ($\circ$), 
30 K ($\ast$), and 300 K ($\blacksquare$); dotted line corresponds to M(H) measured 
for HOPG sample at T = 5 K. Inset depicts M(H) at T = 5 K for MC1 sample on a 
larger field scale.}
\label{fig1}
\end{figure}

Figure \ref{fig1} presents low-field portions of M(H) isotherms recorded for the MC1 sample 
in the field range -50 kOe $\leq$ H $\leq$ 50 kOe and T $\leq$ 300 K, providing 
evidence for the FM hysteresis loops occurrence at all measuring temperatures. 
The inset in Fig. \ref{fig1} exemplifies the M(H) measured at T = 5 K on a larger field scale, 
which demonstrates that M(H) becomes reversible at H $\geq$ 5 kOe, as well as that 
$\mid$M(H)$\mid$ continuously increases with the field increasing, so that 
$M(H) = M_s + \chi H$, where $M_s(5 K) = 0.02~$ emu/g is the spontaneous magnetization 
obtained from the extrapolation of the linear M(H) region to H = 0, and $\chi(5 K) = 5.1 \times 10^{-6}$
 emu/gOe is the paramagnetic susceptibility. Additionally, in Fig. \ref{fig1} we have plotted M(H) obtained for HOPG 
sample \cite{6} at T = 5 K and H applied parallel to the sample basal planes. 
The spontaneous magnetizations M$_{s}^{HOPG}$(5 K) = 0.0007 emu/g obtained for HOPG sample testifies  
that FM is drastically enhanced in the MC as compared to the HOPG. We 
stress that because the MC exhibits a very high surface area, the iron impurities 
should be (1) present as very small particles, and (2) finely dispersed on the 
surface of MC. Then, since the HOPG sample contains $\sim$ 100 ppm of Fe \cite{6}, 
which exceeds the total magnetic impurity contents in our MC, the 
inequality M$_{s}^{MC}$ $\gg$ M$_{s}^{HOPG}$ cannot be accounted for by the 
impurity effect (see also below).

\begin{figure}
\begin{center}
\epsfig{file=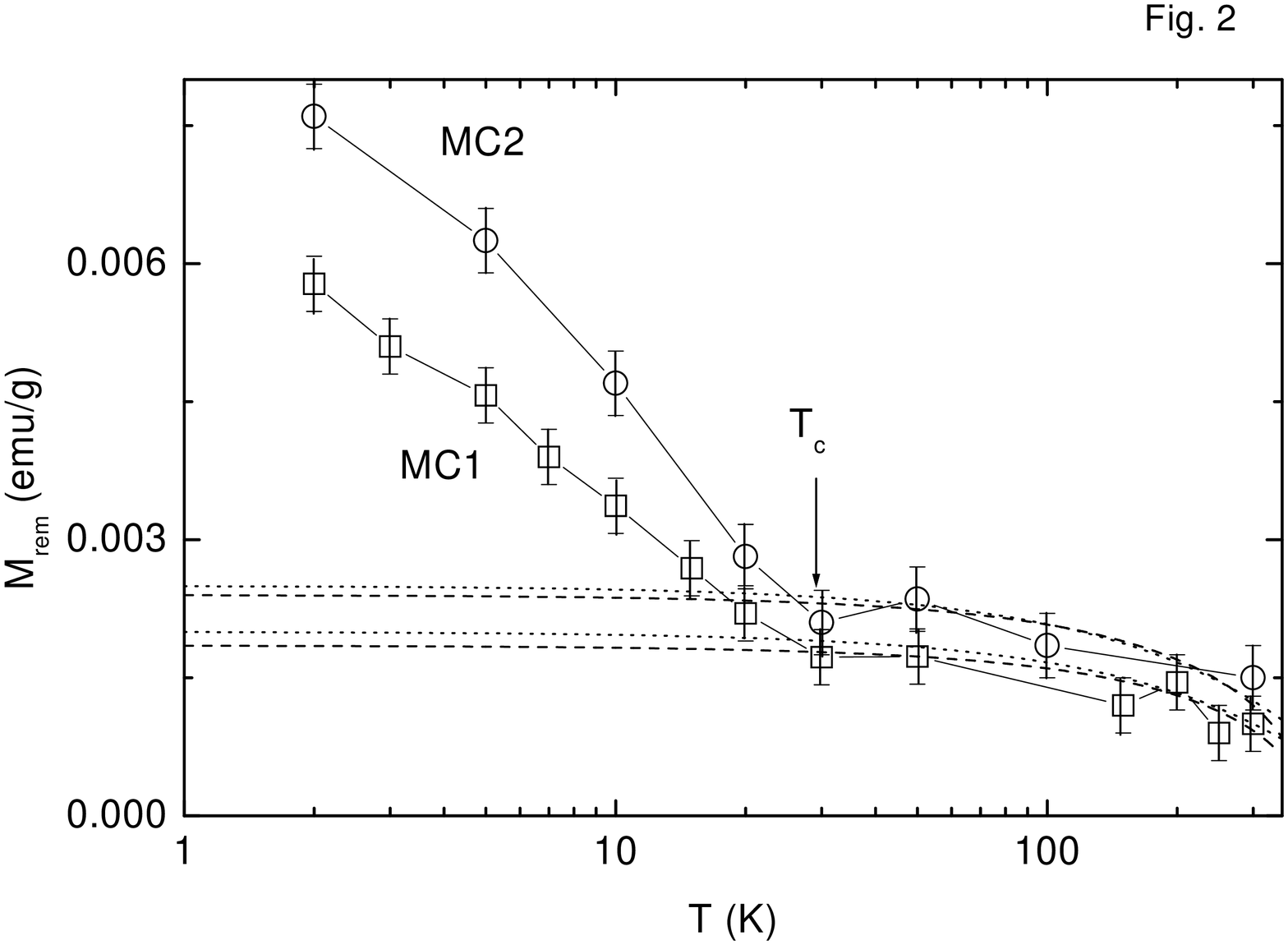,width=\columnwidth}
\end{center}
\vspace{-0.5cm} \caption[*]{Remanent magnetization, M$_{rem}$(T) obtained for MC1 and MC2 samples. 
Dotted and dashed lines correspond to the dependence  M$_{rem}(T) \sim (1 - T/T^{*})^n$ with
$n=1$ (1/2) and T$^{*}$ = 600 (400)K, respectively. Arrow indicates the temperature T$_{c}$ associated 
with the percolative-like FM transition.}
\label{fig2}
\end{figure}

\begin{figure}
\begin{center}
\epsfig{file=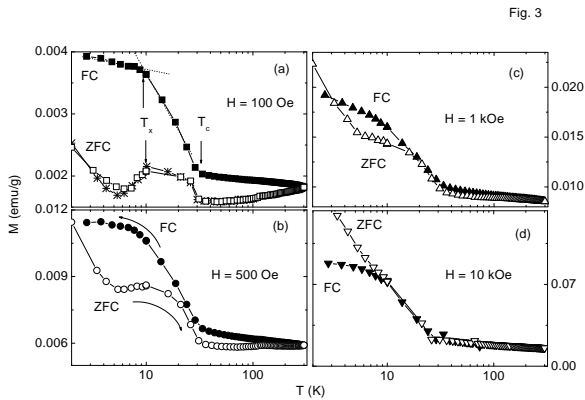,width=\columnwidth}
\end{center}
\vspace{-1cm} \caption[*]{(a - d) M(T, H) measured for MC2 sample in both ZFC (open symbols)  and FC (solid symbols)
regimes at various 
applied magnetic fields; see text for meaning of T$_{c}$ and T$_{\times}$(a), dotted lines are a guide 
for the eye. $M_{ZFC}(T, H = 100 Oe)$ (*) measured independently for $2 K \leq T \leq 100 K$ is included 
in (a) for completeness.}
\label{fig3}
\end{figure}

The remanent magnetization 
M$_{rem}$(H = 0) = [M$^{+}$(H = 0) - M$^{-}$(H = 0)]/2 versus temperature obtained 
for both MC1 and MC2 samples, where M$^{+}$(H = 0) and M$^{-}$(H = 0) are zero-field 
positive and negative magnetizations measured after the field cycling is shown in Fig. \ref{fig2}. As Fig. \ref{fig2} 
demonstrates, the M$_{rem}$(T) steeply increases at T $\leq$ T$_{c}$ $\approx$ 30 K. The 
steep increase of the magnetization below 30 K can also be seen in Fig. \ref{fig3} (a - d) 
where M(H,T) measured for the MC2 sample in both zero-field-cooled (ZFC) and field-cooled 
on cooling (FC) regimes at various applied fields are given; the magnetization data 
corresponding to the ZFC regime, M$_{ZFC}$(T,H), were taken on heating after the sample cooling 
at H = 0, and the magnetization in the FC regime, M$_{FC}$(T,H), was measured as a function of 
decreasing temperature in the applied field. The difference between M$_{ZFC}$(T,H) and M$_{FC}$(T,H) 
magnetizations, which is apparent from Fig. \ref{fig3}, suggests that the MC possesses a
disordered magnetism similar as, for instance, diluted FM semiconductors 
(DFMS) \cite{17,18}, doped LaMnO$_{3}$ (``manganites'') \cite{19}, and TDAE-C$_{60}$ \cite{20}. 
The salient feature of the data presented in Fig. \ref{fig3} is the negative sign of the 
difference $\Delta M(T, H) = M_{FC}(T, H) - M_{ZFC}(T, H)~$ observed at H = 10 kOe and $T < 10 K$, 
[see Fig. 3 (d)], which is a rare phenomenon occurring, however, in a presence of metastable 
magnetic phases \cite{Ro, 18}.

We proceed with a discussion of the results noting a certain similarity between 
the magnetic behavior of MC and that of DFMS and manganites,
where an occurrence of the percolative-type FM transition has been documented
\cite{17,21,22,23,24,25,26}. According to the 
percolation picture, uncorrelated FM clusters are formed below a temperature T$^{*}$, 
leading to finite although small values of M$_{s}$(T,H), M$_{rem}$(T,H), and 
$\Delta$M(T, H) \cite{17}. As the temperature decreases, FM correlations develop 
on a larger scale, and eventually a long-range FM order emerges at the critical 
temperature T$_{c}$ $<$ ($\ll$) T$^{*}$. The random orientation of M$_{s}$ corresponding 
to different FM clusters can be stabilized in the presence of quenched disorder 
and/or competing AFM correlations leading to the spin-glass-like behavior \cite{27,28}. 
Note, that the magnetization anomaly at $T = T_{\times} \approx 10~$ K, viz., both the 
step in $M_{ZFC}(T)$ and the kink in $M_{FC}(T)$, see Fig. \ref{fig3}(a),
is consistent with the occurrence of AFM correlations competing 
with the FM order \cite{new29}. The applied magnetic field aligns FM clusters diminishing 
$\Delta$M(T, H), as experimentally found ($H>500 Oe~$) (Fig. \ref{fig3}). However, 
$\Delta$M(T, H) $<$ 0 observed at high enough fields and T $<$ T$_{c}$, see 
Fig. \ref{fig3} (d), may be associated with coexisting metastable AFM and FM phases, 
as well as with a first order nature of the percolative transition \cite{26}. 

Adopting here the percolative-type picture, we tend to identify T$_{c}$ $\approx$ 30 K 
with a transition temperature below which an enlargement of preexisting FM clusters 
takes place. At the same time, because the experimental data indicate a coexistence 
of different magnetic states, the T$_{c}$ should not be associated with a global FM phase 
transition temperature. Furthermore, as Fig. \ref{fig2} shows, M$_{rem}$(T) can be well 
approximated by the dependence M$_{rem}(T) \sim (1 - T/T^{*})^n$ at T $\geq$ 
T$_{c}$, where n = 1 (1/2) and T$^{*}$ = 600 (400) K can be associate with the temperature below which isolated FM 
clusters are formed. Note, that the linear $M_{rem}$ vs. T dependence has been observed for 
HOPG \cite{6,7} and graphite-sulfur composites \cite{Sergio} in a broad temperature range.

Apparently, the above phenomenology describes well the experimental results. At the same 
time, the observation of curved graphene sheets in our MC, presumably containing pentagons 
and heptagons \cite{15}, is consistent with theoretical expectations of the occurrence 
of both FM and AFM instabilities in topologically disordered graphitic layers \cite{14}; 
at low energies, density of states of Dirac fermions diverges as 
N(E) $\sim$ E$^{-1+2/z}$ (the dynamical exponent z depends on disorder) or 
N(E) $\sim$ E$^{-1}$ exp(- $\mid$ lnE $^{2/3}$ $\mid$) \cite{29,30}, leading to the observed 
electronic instabilites. It is also possible that a peak in the N(E) at E = 0 \cite{31} 
triggers further the instabilities. Since HOPG has a small number (if any) of topological 
defects \cite{14}, the experimental observation that M$_{s}$$^{HOPG}$ $\ll$ M$_{s}$$^{MC}$ 
(see above) is not unexpected. On the other hand, there exists both experimental and 
theoretical evidence that fullerene-like fragments with positive and/or negative curvature 
should be a common feature of microporous carbons \cite{32}. Such fragments should naturally 
appear also in rh-C$_{60}$ samples with partially destroyed C$_{60}$ molecules \cite{33,9,10,11}. 

\begin{figure}
\vspace{0.5cm}
\begin{center}
\epsfig{file=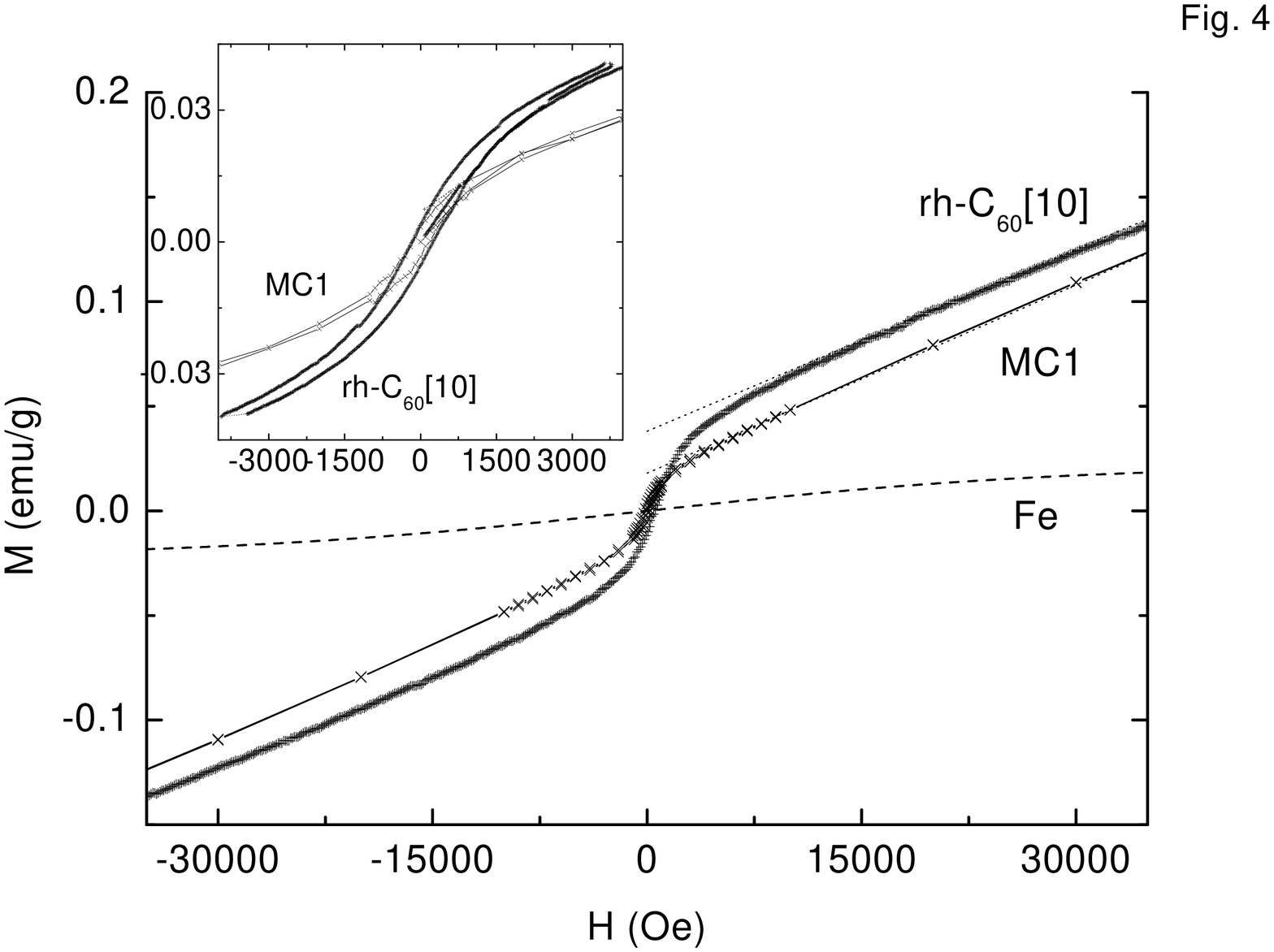,width=\columnwidth}
\end{center}
\vspace{-1cm} \caption[*]{M(H) measured at T = 10 K for rh-C$_{60}$ \cite{10} and MC1 samples;
dotted lines are obtained from the equation M(H) = M$_{s}$ + $\chi$H, where M$_{s}$ = 0.018 emu/g (MC1),
M$_{s}$ = 0.038 emu/g (rh-C$_{60}$), and $\chi$ $\approx$ 3 $\cdot$ 10$^{-6}$ emu/g $\cdot$ Oe for both
rh-C$_{60}$ and MC1 samples; inset demonstrates the M(H) at low fields. Dashed line corresponds to the
calculated magnetization $M_{Fe}(H, T = 10 K)$ for Fe (64 ppm) impurities using a Brillouin function, see text.}
\label{fig4}
\end{figure}

A comparative analysis of the data obtained on MC1 and rh-C$_{60}$ sample synthesized at the 
pressure 9 GPa and T = 800 K \cite{10}  (Fig. \ref{fig4}) reveals a striking correspondence 
between M(H) measured in these materials. As follows from Fig. \ref{fig4}, the high field 
portion of the magnetization curves measured at T = 10 K can be fitted by the equation 
M(H) = M$_{s}$ + $\chi$H, where M$_{s}$ = 0.018 emu/g (MC1), M$_{s}$ = 0.038 emu/g (rh-C$_{60}$), 
and $\chi$ $\approx$ 3 $\times$ 10$^{-6}$ emu/gOe is the paramagnetic susceptibility 
for both materials. The remanent magnetizations measured in MC1, $M_{rem} = 0.0034 emu/g$, 
and rh-C$_{60}$, $M_{rem} = 0.0047 emu/g$ samples, see inset in Fig. \ref{fig4}, also 
practically coincide. To be more specific, we included in Fig. \ref{fig4} the calculated 
magnetization due to Fe impurities (64 ppm) $M_{Fe}(H) = N\mu \tanh(\mu H/k_BT)$, where 
$T = 10~$ K, $N$ is the number of Fe ions per gram, and $\mu = gJ\mu_B$, taking the 
Lande factor $g = 2$ and the spin $J = 2.5$ for $Fe^{3+}$. It is evident from Fig. \ref{fig4} 
that magnetic impurities cannot account for the measured magnetization.

On the other hand, larger values of M$_{s}$(10 K) $\approx$ 0.2 $\ldots$ 0.4 emu/g and 
``saturated'' character of FM hysteresis loops were reported for rh-C$_{60}$ samples synthesized 
at the pressure of 6 GPa and T = 1020 $\ldots$ 1050 K \cite{9,11}. This can be understood assuming 
a stronger coupling between FM clusters in 6 GPa-rh-C$_{60}$ samples \cite{9,11} as compared to 
that in MC and 9 GPa-rh-C$_{60}$ compounds \cite{10}, in a close analogy with the magnetization 
behavior in other inhomogeneous ferromagnets \cite{17,23}. Actually, our recent measurements performed on 
6 GPa-rh-C$_{60}$ samples synthesized at T = 1073 $\ldots$ 1123 K, revealed a magnetic behavior 
characteristic of inhomogeneous ferromagnets \cite{34}, suggesting that the sample (rh-C$_{60}$) preparation 
conditions control an interaction strength between FM clusters and/or their size.

To summarize, we reported on the observation of high-temperature ferromagnetism in microporous carbon 
consisting of curved graphitic layers. The results provide evidence that FM behavior of MC is governed 
by isolated clusters in a broad temperature range, and suggest the occurrence of percolative-type 
transition with the temperature lowering. We pointed out that ferromagnetism in both MC and rh-C$_{60}$ compounds 
can be associated with fullerene-like fragments with positive and/or negative curvature.

This work was partially supported by FAPESP and CNPq. We acknowledge helpful discussions with Prof. A. 
Tomita (Institure of Multidisciplinaty Research for Advanced Materials, Tohoku University, Japan).


\begin{thebibliography}{99}
\bibitem{12} The M$_{s}$ comparable to that of magnetite has been reported in Ref. \cite{2} 
and Ref. \cite{8}. As emphasized, however, by authors of Ref. \cite{8}, the anomalously 
large M$_{s}$ can be not intrinsic to carbon but result from a proximity effect due 
to an extrinsic magnetic material inclusions.
\end{thebibliography}
\end{document}